\documentclass[prb,aps,twocolumn,showpacs,superscriptaddress,amsmath,amsfonts]{revtex4-2}
\usepackage{graphicx,color}
\usepackage[english]{babel}
\usepackage{amsmath,amssymb}
\usepackage{stmaryrd,dsfont,txfonts}
\usepackage{subfigure}
\usepackage{mathrsfs}
\usepackage{colordvi}
\usepackage{epsfig}
\usepackage{float}
\usepackage{dcolumn}
\usepackage{ifpdf}
\usepackage{siunitx}
\usepackage{physics}
\usepackage{nicematrix}
\usepackage{hyperref}
\hypersetup{
	colorlinks,%
	citecolor=blue,%
	filecolor=blue,%
	linkcolor=blue,%
	urlcolor=blue
}

\newcommand\identity{1\kern-0.25em\text{l}}

\begin{document}
\title{Multiband $k \cdot p$ theory for hexagonal germanium}
\author{Yetkin Pulcu}
\email{yetkin.pulcu@uni-konstanz.de}
\affiliation{Department of Physics, University of Konstanz, D-78457 Konstanz, Germany}
\author{J{\'a}nos Koltai}
  \affiliation{Department of Biological Physics, E\"otv\"os Lor\'and University, Budapest, Hungary}
\author{Andor Korm{\'a}nyos}
  \affiliation{Department of Physics of Complex Systems, E\"otv\"os Lor\'and University, Budapest, Hungary}
\author{Guido Burkard}
\email{guido.burkard@uni-konstanz.de}
\affiliation{Department of Physics, University of Konstanz, D-78457 Konstanz, Germany}

\begin{abstract}
The direct bandgap found in hexagonal germanium and some of its alloys with silicon allows for an optically active material within the group-IV semiconductor family with various potential technological applications. However, there remain some unanswered questions regarding several aspects of the band structure, including the strength of the electric dipole transitions at the center of the Brillouin zone. Using the $\mathbf{k\cdot p}$ method near the $\Gamma$ point, including 10 bands, and taking spin-orbit coupling into account, we obtain a self-consistent model that produces the correct band curvatures, with previously unknown inverse effective mass parameters, to describe 2H-Ge via fitting to {\it ab initio} data and to calculate effective masses for electrons and holes. To understand the weak dipole coupling between the lowest conduction band and the top valance band, we start from a spinless 12-band model and show that when adding spin-orbit coupling, the lowest conduction band hybridizes with a higher-lying conduction band, which cannot be explained by the spinful 10-band model. With the help of L\"owdin's partitioning, we derive the effective low-energy Hamiltonian for the conduction bands for the possible spin dynamics and nanostructure studies and in a similar manner, we give the best-fit parameters for the  valance-band-only model that can be used in the transport studies. Using the self-consistent 10-band model, we include the effects of a magnetic field and predict the electron and hole \textit{g} factor of the conduction and valance bands. Finally, we give an ellipticity analysis of the found effective mass tensor, to ensure the uniqueness of the solutions for its  application to heterostructures.
\end{abstract}

\maketitle

\section{Introduction}

Optical activity plays a crucial role in semiconductor materials due to the possible optoelectronic integration which is vital for optoelectronics and integrated photonics, optical modulation, and light emission. \cite{soref2006past,sun2016optical,nakamura1998roles}. However, silicon technology cannot be used for these purposes due to the indirect band-gap of cubic Si (3C-Si) although much effort has been made to turn 3C-Si into an efficient emitter \cite{canham2000gaining, green2001efficient}. In recent years, the hexagonal 2H-Ge phase of germanium has peaked in interest due to the its direct bandgap. Experiments by Fadaly \textit{et al.}\ \cite{fadaly2020direct}, showed that hexagonal germanium has a weak but non-zero optical activity. It has  also been demonstrated that the radiative lifetime can be increased by more than three orders of magnitude when a certain percentage of germanium atoms are replaced by silicion, which makes it as optically active as GaAs. Similar results are also obtained theoretically by R\"odl \textit{et al.}\ \cite{fadaly2020direct} by using {\it ab initio} calculations of the radiative lifetime of hex-Ge near the $\Gamma$ point. Interestingly, the radiative lifetime obtained from experiments and {\it ab initio} calculations show a disagreement by almost an order of magnitude that is yet to be explained.  

While Ge and Si share similar chemical properties, their behavior is different in the hexagonal crystal structure. 
Similar to 3C-Si, hexagonal Si (2H-Si) which is described by the lonsdaleite crystal structure also has an indirect bandgap with the lowest conduction band (CB) located at the M point in its Brillouin zone, rather than the X point \cite{fadaly2020direct} as in the case for 3C-Si. In contrast to Si, the transition from cubic (3C-Ge) to lonsdaleite (2H-Ge) germanium is concomitant with the high-symmetry L point along the $[111]$ axis in the cubic phase folding onto the $\Gamma$ point in the hexagonal phase. The folding of the high-symmetry point L  to $\Gamma$ is important as the lowest CB in 3C-Ge is located at the L point, which maps to the $\Gamma$ point, making hex-Ge a direct bandgap semiconductor, with a bandgap of $0.3$ eV \cite{fadaly2020direct}. Belonging to the $P6_3/mmc$ space group, the point group for 2H-Ge at the $\Gamma$ point of the Brillouin zone is $D_{6h}$. This is quite similar to the wurtzite crystal structure with the $C_{6v}$ point group at $\Gamma$, and we can write $D_{6h} = C_{6v} \otimes C_s$ with $C_s = \identity \otimes \sigma_h$ and $\sigma_h$ being the mirror symmetry, such that correctly symmetrized $C_{6v}$ bases can be used for the group $D_{6h}$.

It is well known that $\mathbf{k\cdot p}$ theory \cite{voon2009kp} is a very useful tool for band structure studies. It is applicable not only to bulk materials but can be adapted to describe nanoscale and low-dimensional structures. In the past, $\mathbf{k\cdot p}$ theory has been successfully used for different materials such as cubic Si and Ge \cite{cardona1966energy,richard2003energy,humphreys1981valence}, III-V compounds in the wurtzite  phase \cite{chuang1996k,andreev2000theory,gutsche1967spin, marquardt2020multiband}, monolayers of transition metal dichalcogenides \cite{kormanyos2013monolayer,kormanyos2015k}, and for calculating the Land\'e $g$ factor \cite{roth1959theory,hermann1977k}, Landau levels \cite{junior2019k,shahbazi2022effective}, and strain effects \cite{bahder1990eight,chuang1997band,stier1999electronic}. As the method is based on group-theoretical selection rules \cite{tronc1999optical}, it is a very powerful tool for calculations of optical transition matrix elements, in which we can explain the low optical activity of the lowest CB for 2H-Ge and other possible transitions, as well as the effect of the spin-orbit coupling.  $\mathbf{k\cdot p}$ theory is a semi-empirical method and requires a number of material-specific parameters either from experiments \cite{cardona1966energy,chuang1996k} or {\it ab initio} calculations \cite{dugdale2000direct}. 

In this paper, we derive a $10\times 10$ $\mathbf{k \cdot p}$ Hamiltonian to describe the  band structure of 2H-Ge, in accordance with {\it ab initio} calculations. We show that L\"owdin's formalism must be used to correctly describe the lowest CB in $k_x$ ($\Gamma \rightarrow K$) and highest valence band (VB) in $k_z$ ($\Gamma \rightarrow A$) direction. We also find the best fit values for the optical transition elements (Kane or momentum $\bf{p}$ matrix elements) and Bir-Pikus parameters, similarly to the study of Chuang and Chang (CC) \cite{chuang1996k}. Using these parameters, we obtain the effective masses for five bands (2-CB and 3-VB) via parabolic fit. To understand the weak optical activity of the lowest CB, we develop a $12\times 12$ spinless model and show that the band is only optically active when spin-orbit coupling (SOC) is considered. We also develop low-energy effective models for electrons and holes for the possible usage in the heterostructures and transport properties. Using the momentum matrix elements and energy splittings, we evaluate the $g$ factor for the highest energy VBs and the  second-lowest CB. Finally, to make heterostructure studies of hex-Ge more reliable, we also analyze the ellipticity of the effective mass tensor and perform a new parameter fit to ensure that ellipticity conditions are fulfilled \cite{veprek2007ellipticity,veprek2008operator,veprek2008reliable}.

This paper is organized as follows: In Sec.~\ref{sec:ab-initio}, we begin by presenting the \textit{ab-inito} methods used to parametrize the $\mathbf{k\cdot p}$ model. In Sec.~\ref{sec:kp} we present the multiband $\mathbf{k\cdot p}$ model, together with L\"owdin's formalism. In Sec.~\ref{sec:Fit}, we describe the general fitting procedure to our $10\times10$ model, the fitted optical transition elements, inverse mass parameters and the effective masses for different directions, as well as the weak optical activity of the lowest conduction band with a possible explanation. We also derive the effective low-energy Hamiltonian for the second-lowest conduction band in Sec.~\ref{sec:Low} and compare the findings wit the originally derived $\mathbf{k\cdot p}$ model. The ellipticity of the effective mass tensor, its importance, and a second parameter fit with constrained ellipticity for hex-Ge is discussed in Sec.~\ref{sec:Ellipticity}. In Sec.~\ref{sec:gFac}, using the 10-band Hamiltonian, we investigate the $g$ factor of electrons and holes when a magnetic field is applied to either parallel or perpendicular to the main axis of rotation. The discussion of how the selection rules can be used to determine non-zero elements for the $\mathbf{k\cdot p}$ model can be found in the Appendix ~\ref{appx:selection}. Appendix~\ref{appx:perturbation_Ham} has the definitions for the expressions used in the text. The allowed optical transitions from the VBs to CB or CB+1 can be found Appendix~\ref{appx:optical_def} with the addition of the selection rules with and without SOC. Finally, Appendix~\ref{appx:11x11} has the spinless $12\times12$ $\mathbf{k\cdot p}$ Hamilonian.

\section{{\it ab initio} calculations}
\label{sec:ab-initio}

\begin{figure}
	\includegraphics[scale=0.75]{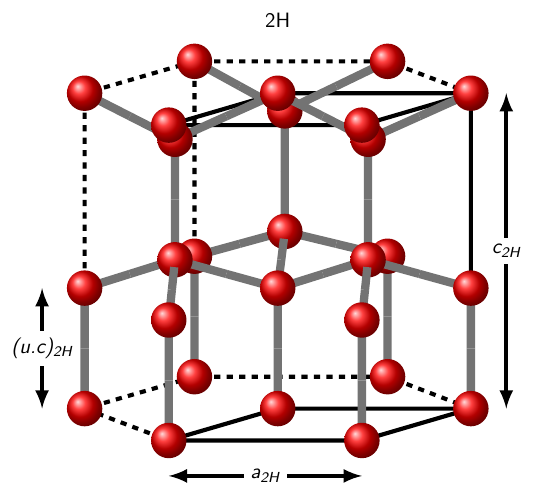}
		\caption{Lonsdaleite or hexagonal diamond (2H) structure where atoms (red balls) are arranged in a hexagonal stacking. The gray lines represent the bonds between atoms. Structural parameters are shown by arrows. Solid black lines indicate the unit cell while the dashed lines visualize the hexagonal structure.}
		\label{fig:FigHex}
\end{figure}

First-principles calculations were carried out by using the Vienna \textit{ab-initio} simulation
package (VASP) \cite{kresse1,kresse2,kresse3} with a plane-wave basis set employed within the framework of the projector augmented-wave method \cite{blochl, kresse4}. 
Geometry relaxation were performed by using PBEsol exchange correlation functionals with a cut-off energy of $500\rm\,eV$, a $12\times 12 \times 6$ Monkhorst-Pack grid sampling of the Brillouin zone and a force criteria of $1\rm\,meV/\AA$. The obtained structural parameters ($a=3.994\,\AA$, $c=6.589\,\AA$ and $u=0.3743$, see Fig.~\ref{fig:FigHex})  are reasonably close to the reported values by Rödl \textit{et al.} \cite{rodl2019accurate}. 
Band structure calculations were performed both with and without the inclusion of the spin-orbit coupling  using the MBJLDA meta-GGA method \cite{mbjlda}. This meta-GGA method is reported to give reliable near-gap energies with a significantly lower calculational cost compared to hybrid functionals (such as HSE06) \cite{rodl2019accurate}.

For the identification of the band symmetries the python tool {\it irRep} was used, which can directly read the Kohn-Sham orbitals of several density functional codes and identifies the irreducible presentation of each bands \cite{irrep1}. If spin-orbit coupling is included, the double crystallographic groups and their representations are incorporated \cite{irrep2}.
We then translated the result of this tool to the notation of \cite{koster1963properties} and found a compelling agreement with the irreducibles published in Ref \cite{rodl2019accurate}.

\section{$\mathbf{k\cdot p}$ Framework}
\label{sec:kp}
The $\mathbf{k\cdot p}$ method has been shown to effectively describe the band structure of the semiconductors around high symmetry points in the Brillouin zone in various studies \cite{willardson1977semiconductors,bastard1986electronic,chuang1996k}. The basic approach is to write the Schrödinger equation in terms of the cell periodic part $u_{n,\bf{k}}(\bf{r})$ of the Bloch wavefunction $e^{i\bf{k}\cdot\bf{r}}u_{n,\bf{k}}$, near the band edge as,
	\begin{align}
	\label{eq:kp}
              Hu_{n,\bf{k}}({\bf{r}}) = \left (H_0 + H_{\rm free} + H_{k\cdot p} + H_{\rm SO}\right)u_{n,\bf{k}}({\bf{r}})=E u_{n,\bf{k}}({\bf{r}}) ,
	\end{align}
where
	\begin{align}
	\begin{split}
	\label{eq:kp_open}
		H_0 &  = \frac{p^2}{2m_0} + V(\bf{r}), \\
		H_{\rm free} & = \frac{\hbar^2k^2}{2m_0}, \\
		H_{k\cdot p} & = \frac{\hbar}{m_0} \bf{k} \cdot \bf{p}, \\
            H_{\rm SO} & = {\frac{\hbar}{4m_{0}^{2}c^{2}}\left[{\boldsymbol{\nabla}}V(\bf{r})\times\bf{p}\right]\cdot\boldsymbol{\sigma}}
	\end{split}
	\end{align}
where $n$ is the band index, ${\bf k}$ is the crystal momentum, $m_0$ is the free electron mass, $V(r)$ is the periodic potential, and $\sigma = (\sigma_x, \sigma_y, \sigma_z )$ are the Pauli spin matrices. The ${\bf k}$-dependent spin-orbit coupling terms are absent throughout this study as they vanish due to the inversion symmetry. Eq.~\eqref{eq:kp} can be solved using  perturbation theory, expanding the $u_{n,\bf{k}}(\bf{r})$ in terms of the known $u_{n,\bf{k}=0}(\bf{r})$, around ${\bf k}=0$.  To obtain the effective band structure using distant band contributions, there are many methods such as folding-down \cite{mccann2013electronic} where one writes a pseudo-Schr\"odinger equation and using norm conserved spinors, obtains a real Schr\"odinger-type equation by performing a series expansion or  L\"owdin's formalism (also known as quasi-degenerate perturbation theory or Schrieffer-Wolff transformation) \cite{lowdin1951note,Schrieffer1966,winkler2003spin}, the method we use in this paper, where the basis functions at the $\Gamma$ point can be divided into the sets A and B. Set A consists of the bands we would like to describe, whereas set B contains all other bands that might give a relevant nonzero contribution to the bands in set A. Using L\"owdin's method and neglecting SOC for now, distant band contributions can be described as,
	\begin{equation}
        \begin{split}
        \label{eq:Lowdin}
            H_{n\times n}({\bf k})_{jj'} =  \sum_{\alpha , \beta} D_{jj'}^{\alpha \beta}k_{\alpha} k_{\beta}, \\
             D_{jj'}^{\alpha \beta} = \frac{\hbar^2}{2m_0} \sum_{\gamma\in B} \frac{p_{j\gamma}^{\alpha} p_{\gamma j'}^{\beta} + p_{j\gamma}^{\beta} p_{\gamma j'}^{\alpha} }{m_0\left (E_0 -E_{\gamma} \right)},
        \end{split}
	\end{equation}
where $(j, j')$ and $\gamma$ belong
to the set A and B, respectively. We should note that set B contributions can only arise as second or higher-order ${\bf k}$-dependent perturbation which can be seen from Eq.~\eqref{eq:Lowdin}.
\begin{table}
	\caption{Basis functions for the $D_{6h}$ point group 
 adapted from CC \cite{chuang1996k}  where $\left\langle {\bf r} | \Gamma_{\alpha}^{n}\right\rangle = u_\alpha^n({\bf r})$. Unlike in CC, there is no $s-p_z$ mixing. $\ket{S_x}$ and $\ket{S_y}$ transform like axial vectors, $\ket{z}$ transforms like a vector and $\ket{\identity}$ transforms like the identity.}
	\begin{ruledtabular}
		\begin{tabular}{cccc}
			& Basis functions & $D_{6h}$ irrep.\  \\ \hline
			$CB+1$ &  $\ket{iz\uparrow}$ & $\Gamma_2^{-}$  \\
			$CB$ & $\ket{\Gamma_3^{-}\uparrow}$ & $\Gamma_3^{-}$ \\
   			$VB$ & $-\frac{1}{\sqrt 2}\ket{S_x+iS_y\uparrow}$  &$\Gamma_5^{+}$ \\
			$VB-1$ & $\frac{1}{\sqrt 2}\ket{S_x-iS_y\uparrow}$ & $\Gamma_5^{+}$ \\
			$VB-2$ &$\ket{\identity\uparrow}$ & $\Gamma_1^{+}$ \\
			$CB+1$ &$\ket{iz\downarrow}$ & $\Gamma_2^{-}$  \\
			$CB$ &$\ket{\Gamma_3^{-}\downarrow}$ & $\Gamma_3^{-}$ \\
   			$VB$ &$\frac{1}{\sqrt 2}\ket{S_x-iS_y\downarrow}$  &$\Gamma_5^{+}$ \\
			$VB-1$ &$-\frac{1}{\sqrt 2}\ket{S_x+iS_y\downarrow}$ & $\Gamma_5^{+}$ \\
			$VB-2$ &$\ket{\identity\downarrow}$ & $\Gamma_1^{+}$ \\
		\end{tabular}
	\end{ruledtabular}
	\label{table:irreps}
\end{table}
\subsection{Five-band model}
\label{subsec:10b-kp}
To describe the effective Hamiltonian of lonsdaleite germanium, we focus on the following five bands: the first conduction band (CB), second conduction band (CB+1), and the first, second, and the third valance bands (VB), (VB-1), and (VB-2), respectively. Including the two-fold spin degeneracy then leads to a total of ten bands. In order to be able to  describe the band structure using $\mathbf{k\cdot p}$ theory, the correct symmetry group and the irreps corresponding to each band have to be known near the point of interest \cite{dresselhaus1955spin, dresselhaus2007group}, that is $\Gamma$ in our case. Previous studies  \cite{rodl2019accurate,de2014electronic} on 2H-Ge have already determined the double-group representation of the bands  at  the   $\Gamma$ point.  However, to effectively use the $\mathbf{k\cdot p}$ method, the relevant single-group representations are needed. Note that, in general different single-group representations can correspond to the same double-group representation. Therefore, we performed {\it ab initio} calculations and obtained the single group basis set, including the two-dimensional spin-1/2 Hilbert space (LS basis), near the $\Gamma$ point (${\bf k}=0$), see Table~\ref{table:irreps}. 

The notation suggests that $u_{n,{\bf k}}({\bf r})$ of the energy band $n$ transforms as the irrep (orbital part) $\left|\Gamma_{\alpha}^{n}\right\rangle$ of the point group $D_{6h}$. Here we follow the K\"oster's notation \cite{koster1963properties} where $S_x$ is an axial vector in the $x$ direction, not to be confused with the projections of the spin up and down ($\uparrow, \downarrow$). Using symmetry considerations of the $D_{6h}$ point group and selection rules (see Appendix~\ref{appx:selection}) with the basis vectors listed in Table~\ref{table:irreps}, we construct the $10\times 10$ Kane-like Hamiltonian to describe the band structure of 2H-Ge near $\Gamma$ point,
\begin{widetext}
	\begin{small}
		\begin{eqnarray}
		\label{eq:HexGeKane}
		H_{\text{Kane}} = \cfrac{\hbar^2 k^2}{2m_0}
		+
		\renewcommand{\arraystretch}{2.0}
		\begin{bmatrix}
		E_{cb+1} & 0 & \frac{-i}{\sqrt{2}}P_2k_{+} & \frac{-i}{\sqrt{2}}P_2k_-  &P_1k_z & 0 & 0 & 0 & 0 & 0\\
		
		0 & E_{cb} & 0 & 0  & 0  & 0 & 0 & 0 & 0 & 0 \\
		
		\frac{i}{\sqrt{2}}P_2 k_- &  0  &  E_v +\Delta_1 +\Delta_2 & 0 & 0 & 0 & 0 & 0 & 0 & 0 \\
		
		\frac{i}{\sqrt{2}}P_2k_+  & 0  & 0 &  E_v +\Delta_1 -\Delta_2   & 0 & 0 & 0 & 0 & 0 &\sqrt{2}i\Delta_3 \\
		
		P_1k_z  & 0  & 0 & 0 & E_v & 0 & 0 & 0 &\sqrt{2}i\Delta_3  & 0 \\
		
		0 & 0 & 0 & 0 & 0 & E_{cb+1}& 0 & \frac{-i}{\sqrt{2}}P_2k_-  & \frac{-i}{\sqrt{2}}P_2k_+  & P_1k_z \\
		
		0 & 0 & 0 & 0 & 0 & 0 &  E_{cb}  & 0 & 0 & 0     \\
		
		0 &0  & 0 & 0 & 0  & \frac{i}{\sqrt{2}}P_2k_+ & 0 &  E_v +\Delta_1 +\Delta_2 & 0 & 0  \\
  
           0 &0  & 0 & 0 & -\sqrt{2}i\Delta_3  & \frac{i}{\sqrt{2}}P_2k_- & 0 &  0 & E_v +\Delta_1 -\Delta_2 & 0  \\
           
           0 &0  & 0 & -\sqrt{2}i\Delta_3 & 0  & P_1k_z & 0 &  0 & 0 & E_v \\
		\end{bmatrix},
		\end{eqnarray}
	\end{small}
\end{widetext}
where $E_{cb+1}$ and $E_{cb}$ are the band energies for for the two lowest conduction bands at $k=0$, $E_v$ is the reference energy, $\Delta_1$ is the crystal field splitting, $\Delta_2$ and $\Delta_3$ are the SOC parameters, $P_{1, 2}$ are the momentum matrix elements and $k_{\pm} = k_x \pm ik_y$ is the crystal momentum. 
Note that $H_{\rm Kane}$ equals $H$ from Eq.~\eqref{eq:kp} restricted to the 10 abovementioned bands.

Due to the presence of inversion symmetry, the nonzero elements of the $\mathbf{k\cdot p}$  Hamiltonian for a lonsdaleite crystal (such as hex-Ge) differ from those in the case of a wurtzite crystal. For instance, the SOC term between the CB and VB, originally neglected by Chuang and Chang \cite{chuang1996k} but later added to the Hamiltonian by Refs.~\cite{fu2008spin,junior2016realistic,fu2020spin}, is zero for hex-Ge due to the inversion symmetry.

It is obvious from Eq.~\eqref{eq:HexGeKane} that in the $k_z$ direction, the top valance band does not have a ${\bf k}$-dependent term and its effective mass is the same as the free electron mass ($m_0$); hence, being a valence band, it has the wrong curvature if the contributions from the other bands are omitted. Also, the lowest conduction band couples neither via the $\mathbf{k\cdot p}$  nor the SOC term to any other band in set A, where bands in set A are shown in Table~\ref{table:irreps} and all other bands are in set B. Hence, the effective mass along the $k_z$ direction would be the  same as the in-plane effective mass but this is contradictory to {\it ab initio} results of Ref.~\cite{rodl2019accurate}. Overall, we find that the model cannot produce the curvature of the energy bands correctly and in this sense it is not self-consistent. To fix this problem, we use L\"owdin partitioning as explained in Eq.~\eqref{eq:Lowdin} to find the contributions from the bands in set B. Hence, using the Kane-like Hamiltonian with the distant band contributions, the Hamiltonian for which we will be using for the fitting becomes
	\begin{align}
	\label{eq:kp2}
          H_{10\times10}^{\rm hex-Ge} = H_{\text{Kane}} + H_{k \cdot p}^{(2)}
        \end{align}
where $H_{k \cdot p}^{(2)}$ is the Hamiltonian of the distant band contributions  and its definition can be found in Eq.~\eqref{eq:kp2corrected}. We should note that, we disregard the renormalized spin-orbit interaction from the bands in set B to set A.

\subsection{Optical activity of the lowest CBs}

From the discussion in Sec.~\ref{sec:kp} it follows that the CB does not have nonvanishing dipole matrix elements with the VB and, hence, for the $10\times 10$ model, the CB appears to be optically dark. However, as it can be seen from the Table~\ref{table:CB}, when spin-orbit coupling is turned on, the optical transition becomes allowed if the polarization of the exciting light is perpendicular to the out-of-plane rotation axis. 
Indeed, it was found in the DFT calculations of 
Ref.~\cite{rodl2019accurate}, that a weak optical transition does exist between the (top) VB and the (lowest) CB. According to 
Ref.~\cite{rodl2019accurate}, this transition was two orders of magnitude weaker than other optical transitions in the system.

In order to understand the origin and weakness of this optical transition, we investigate $12\times 12$ spinless $\mathbf{k\cdot p}$ model that is presented in Appendix \ref{appx:11x11}, Table~\ref{table:12x12}. Looking at the table, we see that $\left\langle CB+5 \left| k_+p_- + k_-p_+ \right| VB\right\rangle = \gamma_3k_-$. Similarly, the SOC matrix element between CB+5 and CB is nonzero because CB+5 transforms as $\Gamma_6^{-}$, CB as $\Gamma_3^{-}$, and the SOC term in the Hamiltonian as $\Gamma_5^{+}$. Using the selection rule, we then find $\Gamma_6^- \otimes \Gamma_5^+ \otimes \Gamma_3^- \supset \Gamma_1^+$. Hence, it is now plausible to conclude that CB, when the SOC is turned on, consists of a linear combination of the irreps $\Gamma_3^{-}$ (CB) and $\Gamma_6^{-}$ (CB+5), which explains the very weak dipole transition from VB to CB found in Ref.~\cite{rodl2019accurate}. We should also point out that in the double-group representation, we have $\Gamma_3^{-} \otimes \Gamma_7^{+} =\Gamma_8^{-}$ and $\Gamma_6^{-} \otimes \Gamma_7^{+} =\Gamma_8^{-} \oplus \Gamma_7^{-}$ where $\Gamma_7^{+}$ is the double group representation of the spinor. Hence, when SOC is considered, both bands belong to the same double group, which makes the hybridization argument more plausible.  Similar arguments can be used for the transitions to the CB+1 band. From Table~\ref{table:CB+1}, one can check that  C transitions (VB-2 $\rightarrow$ CB+1)  in the $x$-$y$ polarization are only allowed when SOC is considered and hence it is weak compared to A (VB $\rightarrow$ CB+1) and B (VB-1 $\rightarrow$ CB+1) transitions where the same transition is allowed even the SOC is turned off. These arguments are consistent with the dipole matrix elements presented in Ref.~\cite{rodl2019accurate}.

\section{Numerical Fitting Procedure of the $\mathbf{k\cdot p}$ Hamiltonian}
\label{sec:Fit}
Although the $\mathbf{k\cdot p}$ method is effective for the description of the coupling of the bands, it still requires an input either from {\it ab initio} calculations \cite{dugdale2000direct,bastos2016stability} or experiments \cite{cardona1966energy,chuang1996k,chuang1997band} as the applied group-theoretic derivation cannot provide numerical values for material-specific  non-zero parameters. To find the best-fit parameters of the Eq.~\eqref{eq:HexGeKane} with the addition of the other band contributions described in Eq.~\eqref{eq:kp2}, we first determine the parameters that appear in ${\bf k}$-independent terms from the {\it ab initio} calculations at $k=0$. Setting $E_{vb} = 0$ , we can read off the conduction band  energies (see Appendix~\ref{appx:perturbation_Ham}) and the crystal field splitting $\Delta_1$ directly from the DFT band structure calculations described in Sec.~\ref{sec:ab-initio} which were performed without SOC. Diagonalizing Eq.~\eqref{eq:HexGeKane} at $k=0$, we obtain for the band-edge energy differences as
\begin{align}
\begin{split}
    \label{eq:SOCsplit}
        {E}_{vb} - {E}_{vb-1} &= \frac{\Delta_{1} +3\Delta_2}{2} - \sqrt{\left ( \frac{\Delta_1-\Delta_2}{2} \right )^2+2\Delta_3^2}, \\
        {E}_{vb} - {E}_{vb-2} &= \frac{\Delta_{1} +3\Delta_2}{2} +  \sqrt{\left ( \frac{\Delta_1-\Delta_2}{2} \right )^2+2\Delta_3^2}.
\end{split}        
\end{align}
where $E_{vb}, E_{vb-1}$ and $E_{vb-2}$ are the band energies of the top three valance bands at $k=0$. Since we have already fixed the value of $\Delta_1$ using the DFT calculations where SOC was switched off, using Eq.~\eqref{eq:SOCsplit}  one can obtain the value of $\Delta_2$ and $\Delta_3$ by making use of DFT calculations where SOC was taken into account.
It is also noted by Ref.~\cite{rodl2019accurate} that cubic approximation ($\Delta_2 = \Delta_3$) does not work for 2H-Ge.

\begin{table}
\caption{Parameters for the Hamiltonian $H_{10\times10}^{\rm hex-Ge}$ for 2H-Ge, composed of $H_{\text{Kane}}$ and $H_{k \cdot p}^{(2)}$ parameters.  The optical transition matrix elements $p_{\perp}, p_{\parallel}$ are related to the parameters $P_1$, $P_2$ appearing in Eq.~\eqref{eq:HexGeKane}  by $P_1 = -i \frac{\hbar}{m_e}\left\langle \Gamma_{2c^-} \left| \hat{p}_{\parallel}\right| \Gamma_{1v+} \right\rangle= -i \frac{\hbar}{m_e}p_\parallel$ and $P_2 = -i \frac{\hbar}{m_e} \left\langle \Gamma_{2c^-} \left| \hat{p}_{\perp} \right| S_y  \right\rangle= -i \frac{\hbar}{m_e} p_\perp$ and they are given in units of $\hbar / a_0$.  Inverse effective masses are given in units of $\frac{\hbar^2}{2m_0}$ and energy splittings are given in eV. Here, we define $\hat{p}_\perp=\hat{p}_x + i \hat{p}_y$ and $\hat{p}_\parallel=\hat{p}_z$.
For comparison, we have included the notations 
$\Delta_{\rm cf}$,
$\Delta_{\rm so}^\parallel$, and $\Delta_{\rm so}^\perp$
for the energy splittings used in Ref.~\cite{rodl2019accurate}.}
\begin{center}
{\renewcommand{\arraystretch}{1.2}
\begin{tabular*}{1\columnwidth}{@{\extracolsep{\fill}}ccc}
\hline
\hline
Parameter & $ H_{10\times10}^{\rm hex-Ge}$\tabularnewline
\hline
$\text{Optical transition matrix elements}$ &  \tabularnewline 
$p_{\perp}$ & 0.4829  \tabularnewline
$p_{\parallel}$ & 0.6431  \tabularnewline 
$\text{Conduction band effective parameters}$ &  \tabularnewline
$A_{c2\perp}$ &  4.1565  \tabularnewline 
$A_{c1\perp}$ & 9.5120  \tabularnewline
$A_{c2\parallel}$ & 2.4091  \tabularnewline
$\text{Valance band effective parameters}$ &  \tabularnewline
$A_1$ &  $-4.3636$  \tabularnewline 
$A_2$ & $-2.0833$  \tabularnewline
$A_3$ & 2.4545  \tabularnewline 
$A_4$ & $-2.7504$  \tabularnewline 
$A_5$ & 2.7232  \tabularnewline 
$A_6$ & $-3.5421$ \tabularnewline 
$\text{Energy splittings}$ &  \tabularnewline
$\Delta_1=\Delta_{\rm cf}$ &  0.2688 & \tabularnewline 
$\Delta_2=\Delta_{\rm so}^\parallel/3$ & 0.0934  & \tabularnewline
$\Delta_3=\Delta_{\rm so}^\perp/3$ & 0.0908  & \tabularnewline 
\hline
\hline
\end{tabular*}}
\end{center}
\label{tab:par_full}
\end{table}

\begin{figure}
\hspace*{-1.5cm}  
	\includegraphics[scale=0.42]{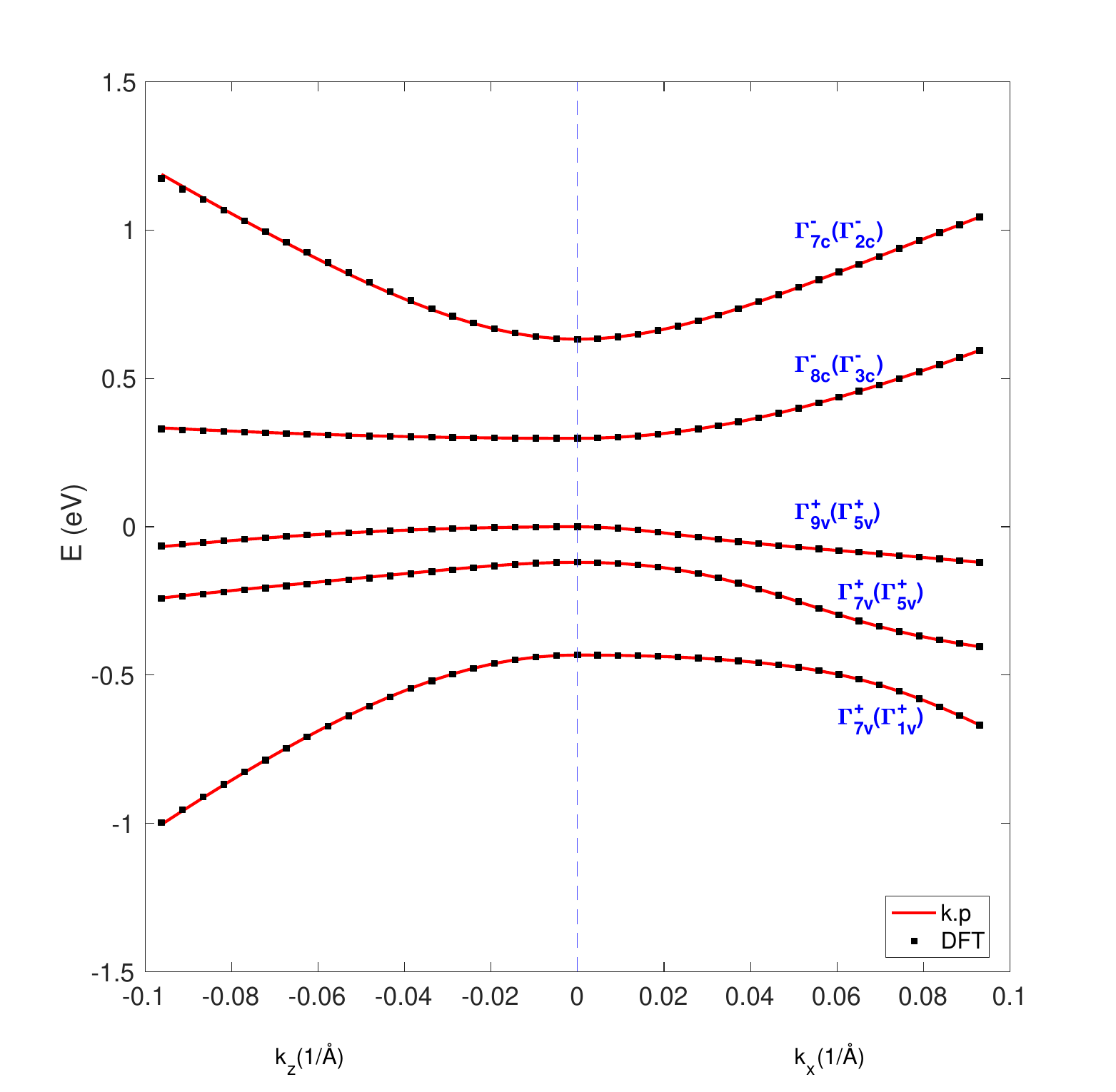}
		\caption{Band structure of hex-Ge around the $\Gamma$ point.
$\mathbf{k\cdot p}$ (solid red lines) fit for the 10 band 2H-Ge, up to the $\SI{0.1}{\angstrom}^{-1}$ to {\it ab initio} (black symbols) for $\Gamma \rightarrow A$ ($k_z$) and $\Gamma \rightarrow K$ ($k_x$) directions. The ordering of the heavy-hole(HH)-light-hole(LH) and crystal-field split-off hole(CH) is due to crystal field splitting being larger than SOC. The combination of inversion and time-reversal symmetry implies that the band structure is doubly degenerate everywhere.}
		\label{fig:FigHam1}
\end{figure}

For the parameters that appear in the ${\bf k}$-dependent terms, we choose our fitting region as $|{\bf k|}\le\SI{0.1}{\angstrom}^{-1}$ due to the limitations of the $\mathbf{k\cdot p}$ method. While fitting, we minimize the objective function $r = \sum_{{\bf{k}}, n} \left [E_{k\cdot p,n}({\bf{k}}) - E_n({\bf{k}})  \right]^2$ where $n$ runs from 1 to 10 and $\bf{k}$ runs along lines from $\Gamma$ to high-symmetry directions up to a cut-off point, and where $E_{k\cdot p,n}({\bf{k}})$ and $E_n({\bf{k}})$ are the fitted and {\it ab initio} values, respectively, which are both plotted in  Fig.~\ref{fig:FigHam1}.

For the calculation of the effective masses, we used a parabolic fit in the vicinity of the $\Gamma$ point; the compiled values can be found in Table~\ref{tab:eff_mass}. The $\Gamma_{8c}^- $ band shows a  highly anisotropic effective mass which is expected as the dipole matrix elements vanish in the $z$ direction but not in the $x$ and $y$ directions. The $\Gamma_{7c}^-$ band is nearly isotropic whereas all the valance bands are anisotropic.

\begin{table}
\caption{The electron and hole masses obtained from the model in Eq.~\eqref{eq:kp2} and,  for comparison, from  DFT calculations (in units of $m_0$). It should be noted that, effective mass in the $x-y$ plane is isotropic.}
\begin{center}
{\renewcommand{\arraystretch}{1.2}
\begin{tabular*}{1\columnwidth}{@{\extracolsep{\fill}}cccc}
\hline
\hline
Bands (Single Irreps) & Direction & $H_{10\times10}^{\rm hex-Ge}$ & DFT (Ref. \cite{rodl2019accurate}) \tabularnewline
\hline
$\Gamma_{9v}^+ \ (\Gamma_{5v}^+)$ & $\Gamma \rightarrow A$  & 0.51 & 0.53 \tabularnewline
  & $\Gamma \rightarrow M$   & 0.08  & 0.07\tabularnewline [1ex] 
  $\Gamma_{7v}^+  \ (\Gamma_{5v}^+) $ & $\Gamma \rightarrow A$  & 0.12 & 0.12\tabularnewline  
  & $\Gamma \rightarrow M$   & 0.10 & 0.10  \\ [1ex] 
    $\Gamma_{7v}^+  \ (\Gamma_{1v}^+)$ & $\Gamma \rightarrow A$  & 0.05 & 0.05 \tabularnewline  
  & $\Gamma \rightarrow M$   & 0.31 & 0.32  \\ [1ex] 
      $\Gamma_{8c}^-  \ (\Gamma_{3c}^-)$ & $\Gamma \rightarrow A$  & 1.00 & 1.09 \tabularnewline  
  & $\Gamma \rightarrow M$   & 0.12 & 0.09  \\ [1ex] 
        $\Gamma_{7c}^- \ (\Gamma_{2c}^-)$ & $\Gamma \rightarrow A$  & 0.04 & 0.04 \tabularnewline  
  & $\Gamma \rightarrow M$   & 0.05 & 0.05  \\ [1ex] 
\hline
\hline
\end{tabular*}}
\end{center}
\label{tab:eff_mass}
\end{table}

\section{Low energy effective models}
\label{sec:Low}
For certain problems, involving $n$ of $p$-doped samples, wires, or quantum dots,  the $10\times 10$ model introduced in Eq.~\eqref{eq:kp2} is not very convenient to use. In this section we provide simpler effective models for the conduction and valence bands separately.  We  compare the effective masses found by Löwdin's partitioning to see if this simpler models can yield comparable results compared to the Hamiltonian we consider in Eq.~\eqref{eq:kp2}. Additionally, we also provide the best fitting parameters for a valance-band-only model.

\begin{figure}
\hspace*{-0.6cm}  
	\includegraphics[scale=0.4]{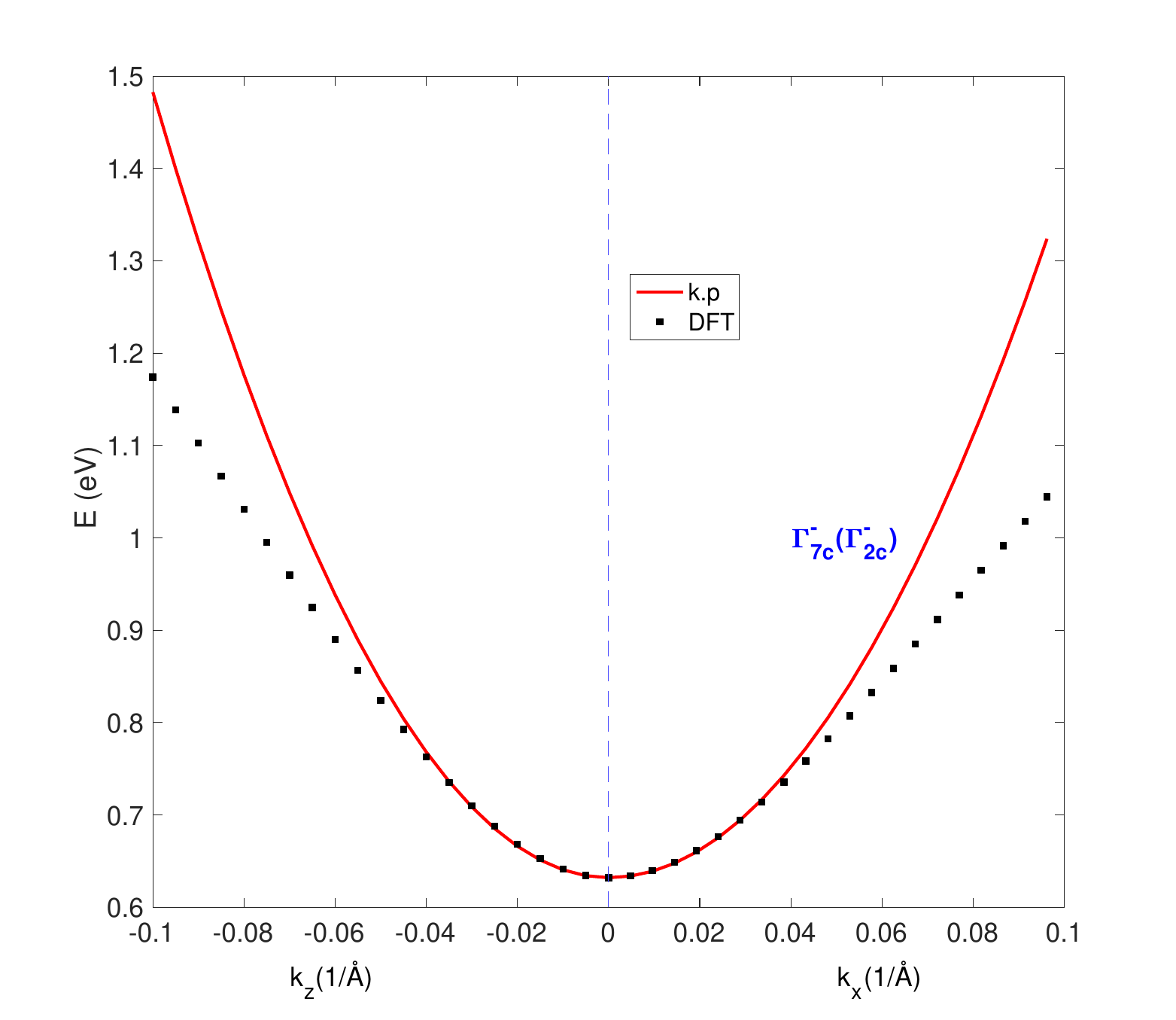}
		\caption{Second-lowest conduction band described by the two-band (spinful CB+1) low-energy effective Hamiltonian for the 2H-Ge. The original fitting region $k\le\SI{0.1}{\angstrom}^{-1}$ has been preserved to show the deviations after   $\SI{0.05}{\angstrom}^{-1}$.}
		\label{fig:cb1}
\end{figure}

We start our discussion with the CB+1 band. One may use the Löwdin's partitioning we have introduced in the Eq.~\eqref{eq:Lowdin} and this time the states that form set A are the $\ket{iz\uparrow}$ and $\ket{iz\downarrow}$ from the Table.~\ref{table:irreps} and all other elements of the table form set B. The effective Hamiltonian can be written as
        \begin{align}
	\label{eq:effective}
             H_{CB+1} = \left [\alpha_1 k_z^2 + \alpha_2\left (k_x^2 + k_y^2 \right)   \right ] \identity_{2\times2},
        \end{align}
with the definitions of $\alpha_1$ and $\alpha_2$ is given by
        \begin{align}
        \label{eq:effective_def}
        \begin{split}
              \alpha_1 &= A_{c2\parallel} + \frac{P_1^2}{E_c - E_v}, \\
              \alpha_2 &= A_{c2\perp} + \frac{1}{2}\frac{P_2^2}{E_c - E_v - \Delta_1 - \Delta_2} + \frac{1}{2}\frac{P_2^2}{E_c - E_v - \Delta_1 + \Delta_2}. 
        \end{split}      
        \end{align}
Here we ignored the third-order corrections that are second-order in ${\bf k}$ and linear in the SOC. Although there are nonzero terms between the spin-up and spin-down channels, they sum up to zero using all the bands that form the set B. From Eq.~\eqref{eq:effective} and using Table~\ref{tab:par_full} for the values that appear in the Eq.~\eqref{eq:effective_def} , we can calculate the effective mass of the CB+1 band in the $k_z$ and $k_x$ directions. In the $\Gamma \rightarrow A$ direction, the effective mass of the electron is $m_{cb+1}^{\perp}/m_0 = 0.04$ and in the $\Gamma \rightarrow M$ and $\Gamma \rightarrow K$ directions  it is $m_{cb+1}^{\parallel}/m_0 = 0.05$. These values are in very good agreement with the values that are found from 10-band fit. From Fig.~\ref{fig:cb1}, it can be seen that the one-band model fits well  to {\it ab initio} up to $\SI{0.05}{\angstrom}^{-1}$ in each direction. We can conclude that the minimal model derived in Eq.~\eqref{eq:effective} can indeed give correctly  that the effective masses of the CB+1 are  slightly anisotropic.

Regarding the CB,  it  does not couple to any other bands in Eq.~\eqref{eq:HexGeKane} and from Eq.~\eqref{eq:kp2}, there is no distant-band contributions to the CB in the $k_z$ direction. Therefore, the effective mass $m_{cb}^{\perp}$ of the CB is very close to the free electron mass in this direction (see Table~\ref{tab:eff_mass}). However, one can check that there is a deep-lying valance band (namely $\Gamma_6^+$ from Table~\ref{table:12x12} in Appendix~\ref{appx:11x11}), which  couples to CB in the $k_x-k_y$ plane. Due to this coupling, the effective mass $m_{cb}^{\parallel}$ of the electron in the $\Gamma\rightarrow M$ and   $\Gamma\rightarrow K$ directions is different from the free-electron mass, see Table \ref{tab:eff_mass}. This means that   the dispersion of the CB is highly anisotropic. The effective Hamiltonian for the CB can be written is the same general form as Eq.~\eqref{eq:effective}, with effective masses $m_{cb}^{\perp}$  and $m_{cb}^{\parallel}$ given in Table \ref{tab:eff_mass}. 

\begin{figure}
\hspace*{-1.2cm}  
\includegraphics[scale=0.4]{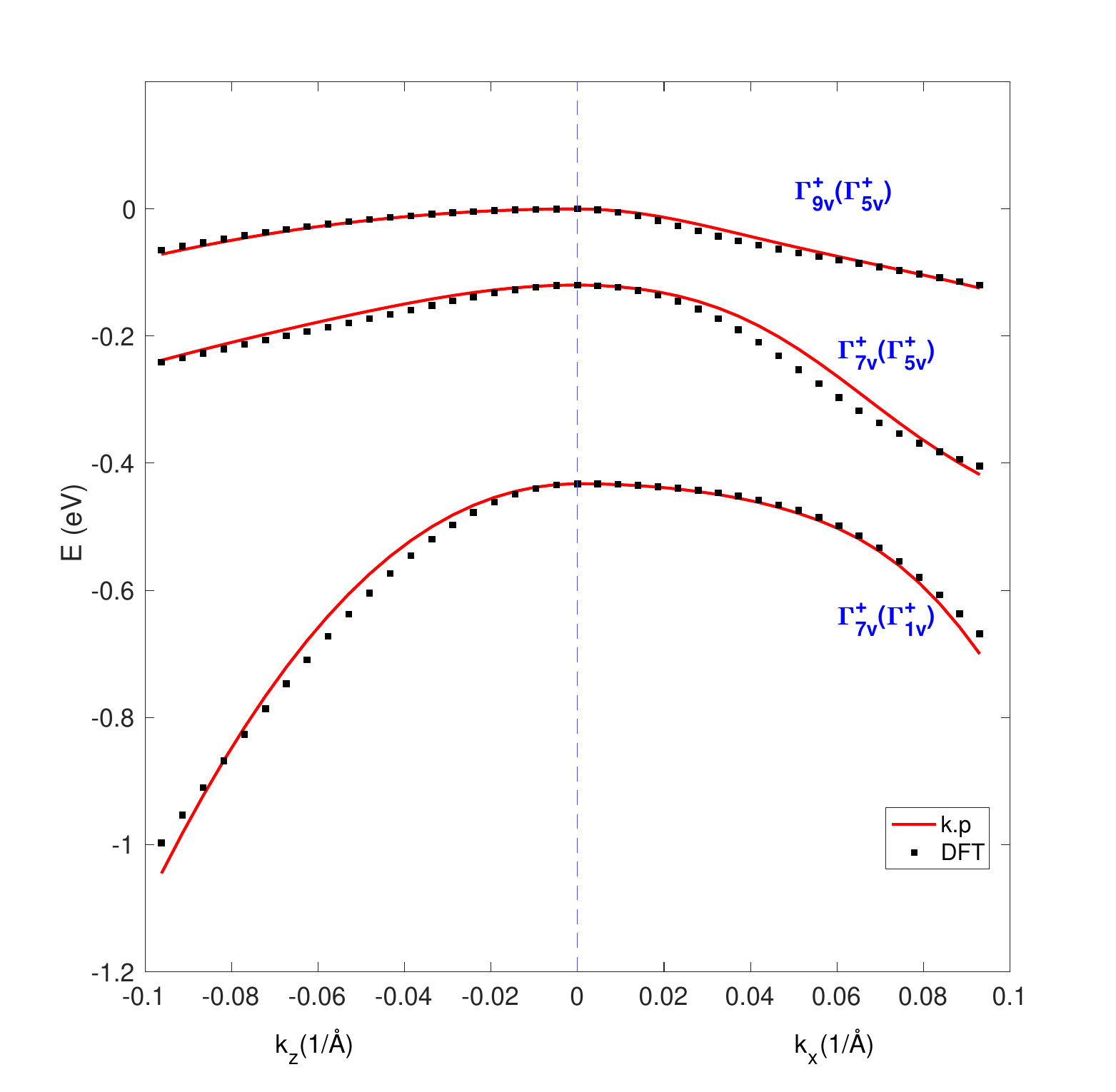}
		\caption{Valence band structure of hex-Ge derived from the six-band (spinful VBs) fit. The original fitting region $k\le \SI{0.1}{\angstrom}^{-1}$ has been preserved to show the deviations, especially for the lower-lying $\Gamma_{1v}^+$ valance band.}
		\label{fig:vbs}
\end{figure}

Finally,  we also give a $6\times6$ fit of the valance-band-only model ($ H_{6\times6}^{\rm hex-Ge}$). We add both conduction bands from Eq.~\eqref{eq:kp2} as distant band contributions to the $6\times6$ valance band only Hamiltonian. To obtain the best-fitting parameters, we use the same approach from Sec.~\ref{sec:Fit}.  We use the energy splittings from Table~\ref{tab:par_full}  and only inverse mass parameters ($A_1, A_2, ...$) are used in the fitting. As it can be seen from the Fig.~\ref{fig:vbs}, there are certain deviations from the {\it ab initio} data, especially for the $\Gamma_{1v}^+$ band in the $k_z$ direction, but the overall agreement is still very good.
\begin{table}
\caption{Parameter sets of the Hamiltonian $ H_{6\times6}^{\rm hex-Ge}$ for 2H-Ge. The units and energy splittings are same as for Table~\ref{tab:par_full}.}
\begin{center}
{\renewcommand{\arraystretch}{1.2}
\begin{tabular*}{1\columnwidth}{@{\extracolsep{\fill}}ccc}
\hline
\hline
Parameter & $H_{6\times6}^{\rm hex-Ge}$\tabularnewline
\hline
$\text{Valance band effective parameters}$ &  \tabularnewline
$A_1$ &  $-17.1282$ \tabularnewline 
$A_2$ & 14.8718  \tabularnewline
$A_3$ & $-2.6087$  \tabularnewline 
$A_4$ & $-6.6647$ \tabularnewline 
$A_5$ & $-7.1014$  \tabularnewline 
$A_6$ & $-11.1423$  \tabularnewline 
\hline
\hline
\end{tabular*}}
\end{center}
\label{tab:par_valance}
\end{table}

\section{Ellipticity Conditions}
\label{sec:Ellipticity}
Often, $\mathbf{k} \cdot \mathbf{p}$ theory is also used to describe heterostructures, such as  quantum wells \cite{chuang1997band, mireles1999ordered, veprek2007ellipticity}, quantum wires \cite{stier1997modeling, park2004finite} and quantum dots \cite{cusack1996electronic, pryor1997electronic}. Since the translational invariance is broken in the growth direction, the quantum number $k_i$ in the $i$th direction is replaced by a derivative operator. Hence, for  nanostructures, the $\mathbf{k} \cdot \mathbf{p}$ Hamiltonian can be written as,
        \begin{align}
        \begin{split}
            	\label{eq:nano}
H_{\mathbf{k} \cdot \mathbf{p}} &= - \sum_{i,j}\partial_i H_{ij}^{(2)}\left(x, k_t \right )\partial_j + \sum_i \left (H_i^{(1)} \left (x,k_t \right )\partial_i + \partial_i(H_i^{(1)} \left (x,k_t \right )\right )\\
&+ H^0\left(x,k_t \right ),
        \end{split}
        \end{align}
where $k_t$ indicates  the direction where there is no confinement. In the early studies, e.g., Ref.~\cite{chuang1997band}, the Hamiltonian is written in a way that terms like, e.g.,  $k_yA_5 k_x$ and $k_x A_5 k_y$ have equal contribution such that $A_5 k_x k_y$ is replaced by the symmetrized form $\frac{1}{2} \left (k_x A_5 k_y + k_y A_5 k_x \right)$ in the heterostructure limit. However, it was later shown  that such an \textit{ad hoc} fix is not guaranteed to give reliable results, and an alternative operator ordering called Burt-Foreman ordering was proposed instead \cite{burt1988new, foreman1993effective}. This means that terms like  $A_5k_xk_y$ are replaced by $k_xA_5^+k_y + k_yA_5^-k_x$, where $A_5 = A_5^+ + A_5^-$ must be satisfied in the bulk limit. Despite its  wide applications, it was shown \cite{white1981electronic} that even Burt-Foreman ordered Hamiltonians might give spurious solutions. To solve this problem, Veprek \textit{et al.} \cite{veprek2007ellipticity,veprek2008reliable,veprek2008operator} proposed a method to find reliable $\mathbf{k} \cdot \mathbf{p}$ parameters in order not to have spurious solutions for heterostructure band energy calculations. It is important to emphasize that the $\mathbf{k}\cdot \mathbf{p}$ parameters  used for the parametrization of the bulk Hamiltonian should be checked for the ellipticity conditions (see, e.g., Ref.~\cite{marquardt2020multiband}) to make the application for heterostructures  more reliable.

To find the ellipticity matrix $h_{ij}^{kl} = \left (H_{ij}^{(2)}\right)_{kl}$, where $k$ and $l$ run over the Bloch-band indices and $i$ and $j$ run over the coordinate axes \cite{veprek2007ellipticity,marquardt2020multiband} , we decouple our $10\times10$ Hamiltonian into $4\times4$ conduction band and $6\times6$ valance band parts. Off-diagonal elements of the $10\times10$ Hamiltonian have no importance in the calculations as they do not contain second derivatives. We can also further simplify the system as spin does not play a role in the ellipticity calculations. Overall, the $9\times9$ (three bands with three directions) ellipticity matrix for the valance band can be written as,
\begin{widetext}
    \begin{equation}
    \label{eq:elliptic}
    h_{ij}^{kl} = 
    \begin{bmatrix}
A_2 + A_4 & -i\left(A_5^+ - A_5^- \right)  &0  &-A_5   & -i\left(A_5^+ + A_5^- \right)   &0  &0  &0  &-A_6^+  \\ 
i\left(A_5^+ - A_5^- \right) &A_2+A_4  &0  &-i\left(A_5^+ + A_5^- \right) & A_5  &0  &0  &0  &-iA_6^+ \\ 
0 &0  &A_1+A_3  &0 & 0  &0  &-A_6^-  &-iA_6^-  &0 \\ 
-A_5 &i\left(A_5^+ + A_5^- \right)  &0  &A_2+A_4  & i\left(A_5^+ - A_5^- \right)  &0  &0  &0  &A_6^+  \\ 
i\left(A_5^+ + A_5^- \right) &A_5  &0  &-i\left(A_5^+ - A_5^- \right)   &  A_2+A_4  &0  &0  &0  &-iA_6^+ \\ 
0 &0  &0  &0  &  0  &A_1+A_3  &A_6^-  &-iA_6^-  &0\\ 
0 &0  &-A_6^-  &0   &   0  &A_6^-  &A_2  &0  &0 \\ 
0 &0  &iA_6^-  &0   & 0  &iA_6^-  &0  &A_2  &0 \\ 
-A_6^+ &iA_6^+  &0  &A_6^+ &  iA_6^+  &0  &0  &0  &A_1\\ 
    \end{bmatrix},
    \end{equation}
\end{widetext}
where $A_{5,6} = A_{5,6}^+ + A_{5,6}^-$. In order for Eq.~\eqref{eq:elliptic} to be elliptic, we demand its eigenvalues $\lambda_n$ to be negative. This condition is equivalent to  demanding that 
\begin{eqnarray}
\label{eq:rho}
    \rho_v = \left|\frac{\sum_{m,{\lambda_m >0}}\lambda_m}{\sum_{n,{\lambda_n <0}}\lambda_n} \right|,
\end{eqnarray}
equals zero. Approximate ellipticity refers to the case where $\rho_v$ is small.  Here  $m$ runs through positive and $n$ through negative eigenvalues. For the conduction band part, using  a similar analysis, we again construct a $6\times6$ ellipticity matrix (two bands with three directions) and in this case we demand that its eigenvalues must be strictly positive. As all the second derivatives for this matrix are on the diagonal and since their coefficients are positive from Table~\ref{tab:par_full}, the eigenvalues are strictly positive already and there is no need to calculate Eq.~\eqref{eq:rho} for the conduction band.
\begin{figure}

\hspace*{-0.5cm}  
\includegraphics[scale=0.45]{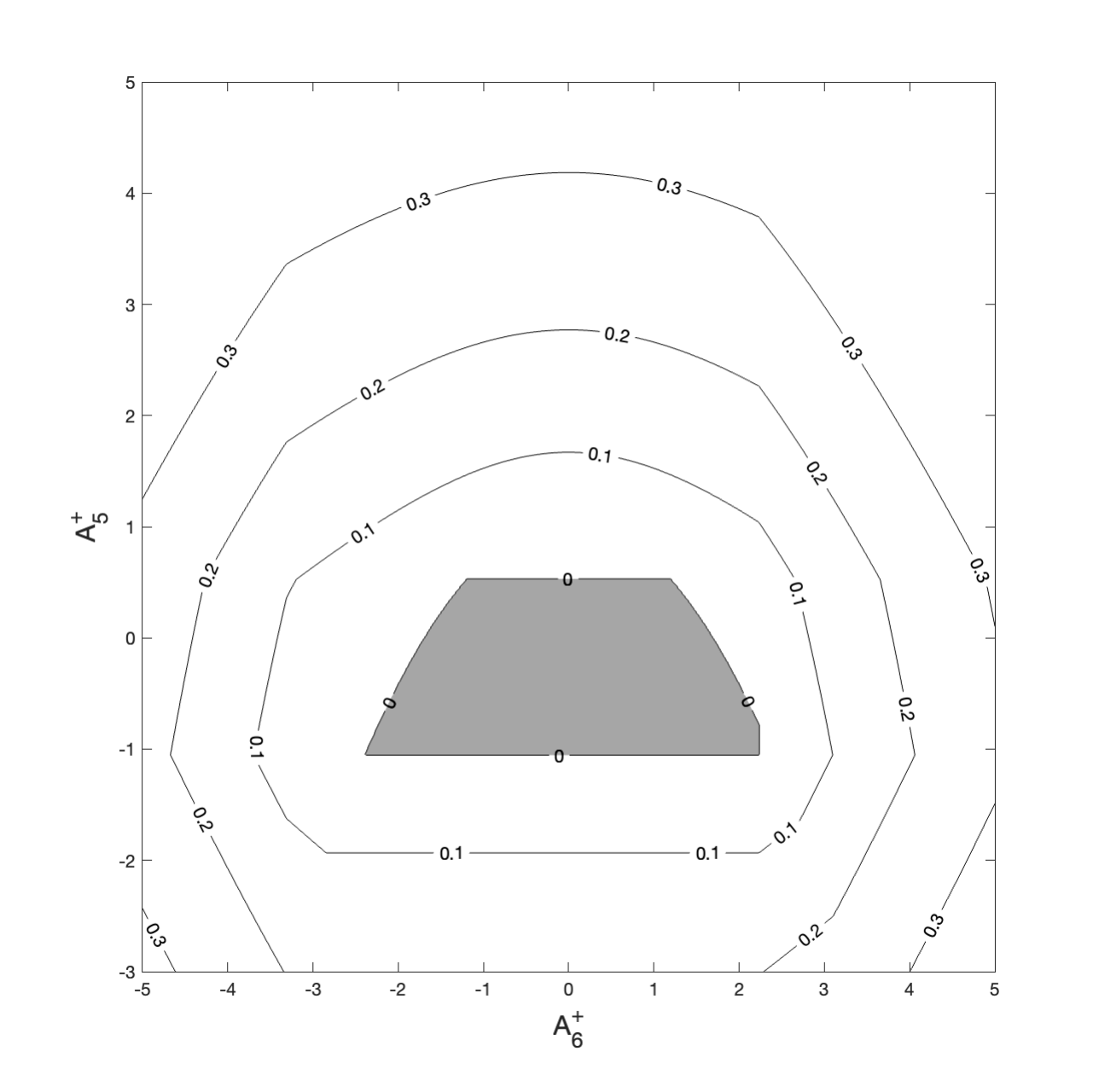}

		\caption{The ratio $\rho_v$ of Eq.~\eqref{eq:rho} for different choices of $A_5^+$ and $A_6^+$ where $A_{5,6} = A_{5,6}^+ + A_{5,6}^-$. The data set from  Table~\ref{tab:elliptic_full} is used in the calculations. The shaded area corresponds to a convex parameter set where Eq.~\eqref{eq:rho} is zero and thus the ellipticity condition is fulfilled.}
		\label{fig:ellips}
\end{figure}

Calculating  $\rho_v$ for the valance band using the parameters from Table~\ref{tab:par_full}, we see that $\rho_v = 0$ cannot be reached no matter how $A_{5,6} = A_{5,6}^+ + A_{5,6}^-$ is split. Hence, it might not be the best choice of parameter set to be used in heterostructure studies although it describes the bulk Hex-Ge perfectly. Therefore, similar to what was done in Sec.~\ref{sec:Fit}, we performed a least square fit to find optimal values for the band energies near the $\Gamma$ point and simultaneously take the ellipticity considerations into account such that $\rho_v = 0$  for a certain $A_5^+$ and $A_6^+$ splitting. The parameters we find are presented in Table~\ref{tab:elliptic_full} and corresponding contour values for the Eq.~\eqref{eq:rho} can be seen in Fig.~\ref{fig:ellips}. Comparing  the values of $A_5$ and $A_6$ in Tables~\ref{tab:par_full}  and \ref{tab:elliptic_full}, one can notice that they differ by roughly an order of magnitude. We have checked that  the dispersion of the bands obtained by using these two parameter sets does not change significantly in the $\mathbf{k}$ range that we consider. This means that  the  $A_5$ and $A_6$ parameters, which correspond to off-diagonal elements in $H^{(2)}_{k\cdot p}$, affect the ellipticity condition much more than the dispersion of the bands.

Similarly, we also calculated  Eq.~\eqref{eq:rho} for the $H_{6\times6}^{\rm hex-Ge}$ model and found that for certain $A_{5,6}^+$ values, $\rho = 0$ can be reached, e.g. the asymmetric splitting where $A_{5,6}^-=0$ and $A_{5,6}^+ = A_{5,6}$. Hence, we do not need an extra parameter set and the parameters from Table~\ref{tab:par_valance} can be used in the heterostructure studies.

\begin{table}
\caption{Elliptic parameters for the $H_{10\times10}^{\rm hex-Ge}$ for 2H-Ge. Optical transition matrix elements and energy splittings are used with their original values. The units follow Table~\ref{tab:par_full}.}
\begin{center}
{\renewcommand{\arraystretch}{1.2}
\begin{tabular*}{1\columnwidth}{@{\extracolsep{\fill}}ccc}
\hline
\hline
Parameter & $ H_{10\times10}^{\rm hex-Ge}$\tabularnewline
\hline
$\text{Optical transition matrix elements}$ &  \tabularnewline 
$p_{\perp}$ & 0.4829  \tabularnewline
$p_{\parallel}$ & 0.6431  \tabularnewline 
$\text{Conduction band effective parameters}$ &  \tabularnewline
$A_{c2\perp}$ &  3.1579  \tabularnewline 
$A_{c1\perp}$ & 9.5120  \tabularnewline
$A_{c2\parallel}$ & 3.3348  \tabularnewline
$\text{Valance band effective parameters}$ &  \tabularnewline
$A_1$ &  $-5.4167$  \tabularnewline 
$A_2$ & $-7.3684$  \tabularnewline
$A_3$ & 3.3328  \tabularnewline 
$A_4$ & 5.5263  \tabularnewline 
$A_5$ & $-0.2631$  \tabularnewline 
$A_6$ & $-0.5415$ \tabularnewline 
$\text{Energy splittings}$ &  \tabularnewline
$\Delta_1=\Delta_{\rm cf}$ &  0.2688 & \tabularnewline 
$\Delta_2=\Delta_{\rm so}^\parallel/3$ & 0.0934  & \tabularnewline
$\Delta_3=\Delta_{\rm so}^\perp/3$ & 0.0908  & \tabularnewline 
\hline
\hline
\end{tabular*}}
\end{center}
\label{tab:elliptic_full}
\end{table}

\section{Effective $g$ Factors}
\label{sec:gFac}
In this section, we investigate Land\'e $g$ factors when a magnetic field is applied either in the $c$ (parallel)  or $x-y$ (perpendicular) crystal axis. If an external magnetic field is applied in the $z$ direction, then the crystal momenta $k_x$ and $k_y$ do not commute, as we can define $\mathbf{k} = \frac{1}{\hbar}\left(\mathbf{p} +e\mathbf{A} \right)$ with the operators $\mathbf{p}$ and $\mathbf{A}(\mathbf{r})$ for the momentum and position-dependent vector potential.  Using the definitions from Eq.~\eqref{eq:Lowdin}, we can split the perturbative terms into a  symmetric and an antisymmetric part as
	\begin{align}
	\label{eq:sym-antisym}
             D_{jj'}^{\alpha \beta}k_{\alpha}k_{\beta} = \frac{1}{2}\left( D_{jj'}^{\alpha \beta} \right)^S \acomm{k_{\alpha}}{k_{\beta}} + \frac{1}{2}\left( D_{jj'}^{\alpha \beta} \right)^A \comm{k_{\alpha}}{k_{\beta}},
        \end{align}
with the definitions,
        \begin{align}
        \begin{split}
    	\label{eq:sym-antisym-def}
             \left( D_{jj'}^{\alpha \beta} \right)^S &= \frac{1}{2} \left [ D_{jj'}^{\alpha \beta} +  D_{j'j}^{\alpha \beta} \right], \\
            \left( D_{jj'}^{\alpha \beta} \right)^A &= \frac{1}{2} \left [ D_{jj'}^{\alpha \beta} -  D_{j'j}^{\alpha \beta} \right].
        \end{split}    
        \end{align}
It has been shown that, from the symmetric part one can obtain the effective mass terms while the Land\'e $g$ factor can be extracted from the anti-symmetric part \cite{roth1959theory,hermann1977k}. Using the double group basis in the form \cite{bayerl2001g},
\begin{align}
\begin{split}
    \label{eqn:DoubleG}
  \Gamma_{7c}^- &=
    \begin{cases}
      \ket{iz\uparrow} \\
      \ket{iz\downarrow} & 
    \end{cases},\\
    \Gamma_{8c}^- &=
    \begin{cases}
      \ket{\Gamma_3^{-}\uparrow} \\
      \ket{\Gamma_3^{-}\downarrow} & 
    \end{cases},\\
     \Gamma_{9v}^+ &=
    \begin{cases}
      \ket{VB\uparrow}\:(m_j = 3/2) \\
      \ket{VB\downarrow} \:(m_j = -3/2) & 
    \end{cases},\\
      \Gamma_{7v_+}^+ &=
    \begin{cases}
      a \ket{VB-1\uparrow} + b \ket{VB-2\downarrow}\:(m_j = 1/2) \\
      b \ket{VB-2\uparrow} + a \ket{VB-1\downarrow}\: (m_j = -1/2) & 
    \end{cases},\\
    \Gamma_{7v_-}^+ &=
    \begin{cases}
      b \ket{VB-1\uparrow} - a \ket{VB-2\downarrow}\:(m_j = 1/2) \\
     -a \ket{VB-2\uparrow} + b \ket{VB-1\downarrow} \:(m_j = -1/2) & 
    \end{cases},
\end{split}
\end{align}
where $a = \sqrt{1 - q_7^2}$ and $b = q_7$ are coupling constants due to the spin mixing of $\Gamma_7$ states with $q_7 = \sqrt{2} \Delta_3 / E_{vb-1} $, we can write the $g$ factor for a magnetic field applied in the $z$ direction as,
        \begin{align}
        \begin{split}
            \label{eqn:g-fac}
            g_z^* = g_0 + \frac{g_0}{im_0} &\sum_{\gamma\neq c} \frac{\left\langle c\uparrow \left| p_{x} \right| \gamma \right\rangle \left\langle \gamma \left| p_{y} \right| c\uparrow \right\rangle}{E_c-E_{\gamma}} \\
            & - \frac{\left\langle c\uparrow \left| p_{y} \right| \gamma \right\rangle \left\langle \gamma \left| p_{x} \right| c\uparrow \right\rangle}{E_c-E_{\gamma}},
        \end{split}
        \end{align}
where $c$ stands for the conduction band, $g_0=2$ is the bare electron $g$ factor, and $\gamma$ belongs to the states from Eq.~\eqref{eqn:DoubleG}, except the conduction band states. Here we should note that, as there are no $\mathbf{k\cdot p}$ terms between the lowest conduction band and the three valance bands that we have considered in Table~\ref{table:irreps}, the $g$ factor of the CB can be taken as $g_0$ in the model introduced in Sec.~\ref{sec:kp}.
Regarding the other bands, e.g., for the $CB+1$ one finds
\begin{align}
\label{g-factorCB1}
    \frac{g_{CB+1,z}^*}{g_0} -1 = \frac{ p_{\perp}^2}{m_0} \left (\frac{-1}{E_{cb+1}-E_{vb}}   +   \frac{1-q_7^2}{E_{cb+1}-E_{vb-1}} +   \frac{q_7^2}{E_{cb+1}-E_{vb-2}}\right).
\end{align}
As can be readily seen from Eqs.~\eqref{eq:SOCsplit} and \eqref{g-factorCB1}, without SOC terms ($\Delta_2$ and $\Delta_3$), $g_{CB+1}$ would be equal to $g_0$. Similar arguments can be made to find the $g$-factor of the holes.

The electron $g$ factor when a magnetic field is applied in the direction perpendicular to the $c$ axis can also be calculated in this way.   The expression for $g_x$  reads
        \begin{align}
        \begin{split}
            \label{eqn:gx}
            g_x^* = g_0 + \frac{g_0}{im_0} &\sum_{\gamma\neq c} \frac{\left\langle c\uparrow \left| p_{y} \right| \gamma \right\rangle \left\langle \gamma \left| p_{z} \right| c\downarrow \right\rangle}{E_c-E_{\gamma}} \\
            & - \frac{\left\langle c\uparrow \left| p_{z} \right| \gamma \right\rangle \left\langle \gamma \left| p_{y} \right| c\downarrow \right\rangle}{E_c-E_{\gamma}}.
        \end{split}
        \end{align}
For the CB+1 band,         
using Eq.~\eqref{eqn:gx} and  the bases specified in Eq.~\eqref{eqn:DoubleG},  $g_x$ is given by
\begin{align}
\label{g-factorx}
   & \frac{g_{CB+1,x}^*}{g_0} -1  \\
   & =  \frac{ p_{\perp}p_{\parallel}}{m_0} \left (\frac{1}{E_{cb+1}-E_{vb-2}}   -   \frac{1}{E_{cb+1}-E_{vb-1}}\right) \sqrt{2q_7^2\left(1-q_7^2 \right)}.\nonumber
\end{align}

In Table~\ref{table:gfactor}, we give the $g$ factors of the electron and holes. The relative $g$ factor difference between electron and holes can be understood via Eq.~\eqref{eqn:g-fac}. For the CB+1 band, the non-zero contributions are coming from VB where the contribution is positive and VB-1, VB-2 where the contribution is negative. Conversely, for the valance bands, the only non-zero contribution is possible via CB+1, which explains the relative difference between electron and holes in this $10\times10$ model. The reason for $g_x=g_0$ for the VB is that due to the spin mixing scheme in Eq.~\eqref{eqn:DoubleG} there is no momentum matrix element in the $z$direction and the right hand side of the  Eq.~\eqref{eqn:gx} vanishes. The anisotropy of the $g$ factor $g_x = g_y \neq g_z$ is expected due to the hexagonal symmetry where the isotropy between $x-y$ and $z$ axis is broken. Table~\ref{table:gfactor} also follows the trend for wurtzite structures where the absolute value of the $g$ factor for holes is greater than for electrons~\cite{junior2019common}. 

\begin{table}[h!]
\caption{Calculated values for the effective g-factors when a magnetic field is applied in the $x$ or $z$ direction using the $10\times10$ $\mathbf{k\cdot p}$ model.}
\begin{center}
{\renewcommand{\arraystretch}{1.2}
\begin{tabular*}{1.0\columnwidth}{@{\extracolsep{\fill}}
llrr}
\hline 
\hline 
  & \multicolumn{2}{c}{$ H_{10\times10}^{\rm hex-Ge}$} &\tabularnewline
  & $g_{x}$ & $g_{z}$\tabularnewline
\hline 
$CB+1$ &  $-2.691$ & $-3.909$ \tabularnewline
$CB$ &   2.0 & 2.0 \tabularnewline
$VB$ & 2.0 & $-18.225$ & \tabularnewline
$VB-1$ &$-13.959$   &9.874  \tabularnewline
$VB-2$ &13.268   &8.442 \tabularnewline
\hline 
\hline 
\end{tabular*}}
\end{center}
\label{table:gfactor}
\end{table}

\section{Conclusions}
In this paper, using {\it ab initio} analysis, we first calculated the irreps of the two lowest conduction and three highest valance bands for the Germanium in the lonsdaleite phase. We developed $10\times10$ $\mathbf{k\cdot p}$ model to fit to the DFT band structure near $\Gamma$. Using the model, we extracted the dipole matrix elements and the inverse effective masses for the conduction and valance bands that fit to the {\it ab initio} band structure up to $\SI{0.1}{\angstrom}^{-1}$. We also calculated the effective masses of electrons and holes in the vicinity of the $\Gamma$ point and find bands with both anisotropic and isotropic masses. As the $10\times10$ $\mathbf{k\cdot p}$ model is not sufficient to explain the optical transition properties of the lowest conduction band, we added more bands to the original model and showed that due to the SOC, lowest conduction band hybridizes with a higher lying band, which gives a weak optical transition when circularly polarized light is used. Using similar arguments, we also explained the transition amplitudes from the top three valance bands to the second lowest conduction band. Using the Hamiltonian we have derived, we calculated the effective $g$ factor of the electrons and holes for a magnetic field along the $c$ and $x-y$ axis of the crystal. Finally, for the future heterostructure calculations, we provided an ellipticity analysis of the fit parameters which is an important tool to obtain correct sub-band energies for the quantum wells, wires, and dots.
In conclusion, we created a $\mathbf{k\cdot p}$ model that captures  important features of the 2H-Ge and we showed that physical parameters like the $g$ factor of the electron and holes can be found using the Hamiltonian at hand. In real samples, the symmetries of the lonsdaleite structure can be broken by various defects, which would affect, among others, the optical selection rules that we obtained. We leave the study of the effects of such symmetry breaking to a future work.

\section{Acknowledgments}
We acknowledge financial support from the ONCHIPS project funded by the European Union’s Horizon Europe research and innovation programme under Grant Agreement No.~101080022.
A. K. and J.~K. acknowledge the support by the Hungarian Scientific Research
Fund (OTKA) Grant No.~K134437 from the source of the National Research, Development and Innovation Fund. This research was supported by the Ministry of Culture and Innovation and the National Research, Development and Innovation Office within the Quantum Information National Laboratory of Hungary (Grant No. 2022-2.1.1-NL-2022-00004).

\appendix
\section{Selection Rules}
\label{appx:selection}

In order to determine the nonzero matrix elements within the $\mathbf{k\cdot p}$ framework, we can first decide whether  products of the form $\Gamma^i \otimes \Gamma^j \otimes \Gamma^k$ contain the identity irrep ($\Gamma_1$). Only in this case, the corresponding matrix element will be  nonzero. Here $\Gamma^{j}$ corresponds to either $\mathbf{p}$ or $\boldsymbol{\nabla}V({\bf r})\times{\bf p}$ and $\Gamma^{i}$ and $\Gamma^{k}$ to irreps of the bands. In short, under any symmetry operation $\hat{G}$,
\begin{align}
        \label{eq:symmetryG}
            \left\langle \psi_{i}^{\alpha} \left| p_{j}^{\beta} \right| \psi_{k}^{\gamma}\right\rangle & = \left\langle \hat{G}\psi_{i}^{\alpha} \left| \hat{G}p_{j}^{\beta} \right| \hat{G}\psi_{k}^{\gamma}\right\rangle.
\end{align}
The relation Eq.~\eqref{eq:symmetryG} can be used to determine whether a matrix element equals zero for symmetry reasons.  Under a rotation operator, the equality becomes $ \left\langle \psi_{i}^{\alpha} \left| p_{j}^{\beta} \right| \psi_{k}^{\gamma}\right\rangle = e^{i\pi r} \left\langle \psi_{i}^{\alpha} \left| p_{j}^{\beta} \right| \psi_{k}^{\gamma}\right\rangle$ where $r$ can take values depending on the rotation symmetry, e.g., $r=1/3$ for a $60^\circ$ rotation. Depending on the value of the exp function, the momentum matrix elements are zero or non-zero. Besides the rotational symmetries, the $D_{6h}$ point group also has mirror and inversion symmetries. For example, under the inversion symmetry $\left\langle \Gamma_{5}^{+} \left| k\cdot p \right| \Gamma_{5}^{+}\right\rangle = -\left\langle \Gamma_{5}^{+} \left| k\cdot p \right| \Gamma_{5}^{+}\right\rangle$ as the coordinates are not paired and hence this matrix element vanishes.
\section{Perturbation Hamiltonians and Definitions}
\label{appx:perturbation_Ham}
In this appendix, we present the matrix form of the Hamiltonians of Eq.~\eqref{eq:kp} and the corresponding parameters.
$H_{10x10}$ in Eq.~\eqref{eq:HexGeKane} can be written as $ \mathcal{H}_0 + \mathcal{H}_{free} + \mathcal{H}_{k\cdot p} + \mathcal{H}_{SO}$, and here we  provide the definitions of the matrix elements.

$\mathcal{H}_0 = \text{diag}\left [E_{cb+1}, E_{cb}, E_v +\Delta_1, E_v +\Delta_1, E_v \right]$ for each spin channel, with $\left\langle \Gamma_{5v^+}^{S_x} \left| \mathcal{H}_0 \right| \Gamma_{5v^+}^{S_x} \right\rangle = \left\langle \Gamma_{5v^+}^{S_y} \left| \mathcal{H}_0 \right| \Gamma_{5v^+}^{S_y} \right\rangle = E_v +\Delta_1$, $\left\langle \Gamma_{2c^-} \left| \mathcal{H}_0 \right| \Gamma_{2c-} \right\rangle = E_{cb+1}$, and $\left\langle \Gamma_{3c^-} \left| \mathcal{H}_0 \right| \Gamma_{3c-} \right\rangle = E_{cb}$.

For the first-order $\mathbf{k\cdot p}$ Hamiltonian we have,
$H_{\mathbf{k}\cdot\mathbf{p}}=\frac{1}{2}\frac{\hbar}{m_e} (k_{+} \hat{p}_{-} + k_{-} \hat{p}_{+}) + \frac{\hbar}{m_e}k_z\hat{p}_{\parallel}$ where $k_{\pm} = k_x \pm ik_y$ with $P_1 = -i \frac{\hbar}{m_e}\left\langle \Gamma_{2c^-} \left| \hat{p}_{\parallel}\right| \Gamma_{1v+} \right\rangle$ and $P_2 = -i \frac{\hbar}{m_e} \left\langle \Gamma_{2c^-} \left| \hat{p}_{\perp} \right| S_y  \right\rangle  = i \frac{\hbar}{m_e} \left\langle \Gamma_{2c^-} \left| \hat{p}_y \right| S_x \right\rangle$.

For the SOC terms, we have 
        \begin{align}
        \begin{split}
            \Delta_{2} &=\frac{i\hbar}{4 m_{0}^{2}c^{2}}\left\langle \Gamma_{5v^+}^{S_x} \left|\frac{\partial V}{\partial x}p_{y}-\frac{\partial V}{\partial y}p_{x}\right| \Gamma_{5v^+}^{S_y} \right\rangle, \\
            \Delta_{3} &= \frac{i\hbar}{4 m_{0}^{2}c^{2}}\left\langle \Gamma_{1v^+} \left|\frac{\partial V}{\partial y}p_{z}-\frac{\partial V}{\partial z}p_{y}\right| \Gamma_{5v^+}^{S_x}   \right\rangle, \\
            &= \frac{i\hbar}{4 m_{0}^{2}c^{2}}\left\langle \Gamma_{1v^+} \left|\frac{\partial V}{\partial z}p_{x}-\frac{\partial V}{\partial x}p_{z}\right| \Gamma_{5v}^{S_y} \right\rangle.    
        \end{split}
        \end{align}
Hence, Eq.~\eqref{eq:HexGeKane} can be written as the sum
        \begin{align}
        \label{A2}
        H_{\text{free}} +
                \begin{bmatrix}
                H_{k \cdot p}^{(1)} & 0\\ 
                0 & H_{k \cdot p}^{(1)}
                \end{bmatrix}   +                
                \begin{bmatrix}
                \Delta_{\text{diag}}& \Delta_{\text{off-diag}}\\ 
                \Delta_{\text{off-diag}}^* & \Delta_{\text{diag}}
                \end{bmatrix}.
        \end{align}
Matrix representation of the distant band contributions $H_{\text{dist.}}$ in the $S_{x}, S_{y}$, and $\identity$ basis can be written as,
\begin{widetext}
	\begin{small}
		\begin{eqnarray}
		\label{eq:kp2inv}
		H_{\text{dist.}} =
		\renewcommand{\arraystretch}{3.0}
		\begin{pmatrix}		
                A_{c2\perp}(k_x^2+k_y^2) + A_{c2\parallel}k_z^2  & 0  &0  &0  &0 \\ 
                0 &A_{c1\perp}(k_x^2+k_y^2) + A_{c1\parallel}k_z^2  &0  &0  &0 \\ 
                0 &0  &L_1k_x^2+M_1k_y^2+M_2k_z^2  &N_1k_xk_y  &N_2k_xk_z \\ 
                0 &0  &N_1k_xk_y  &M_1k_x^2+L_1k_y^2+M_2k_z^2    &N_2k_yk_z \\ 
                0 &0  &N_2k_xk_z  &N_2k_yk_z  & M_3(k_x^2+k_y^2)+L_2k_z^2
		\end{pmatrix}.
		\end{eqnarray}
	\end{small}
with the definitions,
        \begin{equation}
        \begin{split}
        \label{eq:defkp2not} 
            A_{c1\perp} &= \frac{\hbar^2}{2m_0}  \sum_{\gamma}^{B} \frac{2 p_{\Gamma_{3c}\gamma}^xp_{\gamma\Gamma_{3c}}^x}{m_0\left (E_0-E_{\gamma} \right )} ,  A_{c1\parallel} = \frac{\hbar^2}{2 m_0}   \sum_{\gamma}^{B} \frac{2 p_{\Gamma_{3c}\gamma}^zp_{\gamma\Gamma_{3c}}^z}{m_0\left (E_0-E_{\gamma} \right )} , \\
            A_{c2\perp} &= \frac{\hbar^2}{2 m_0}    \sum_{\gamma}^{B} \frac{2 p_{\Gamma_{2c}\gamma}^xp_{\gamma\Gamma_{2c}}^x}{m_0\left (E_0-E_{\gamma} \right )} ,  A_{c2\parallel} = \frac{\hbar^2}{ 2m_0}  \sum_{\gamma}^{B} \frac{2 p_{\Gamma_{2c}\gamma}^zp_{\gamma\Gamma_{2c}}^z}{m_0\left (E_0-E_{\gamma} \right )} , \\
            L_1 &= \frac{\hbar^2}{2 m_0}\sum_{\gamma}^{B}  \frac{2 p_{S_x\gamma}^x p_{\gamma S_x}^x }{m_o \left (E_0-E_{\gamma} \right )}, L_2 = \frac{\hbar^2}{2 m_0} \sum_{\gamma}^{B}  \frac{2 p_{1\gamma}^z p_{\gamma 1}^z }{m_0 \left (E_0-E_{\gamma} \right )}, \\
            M_1 &= \frac{\hbar^2}{2 m_0}\sum_{\gamma}^{B}  \frac{2p_{S_x\gamma}^y p_{\gamma S_x}^y }{m_0 \left (E_0-E_{\gamma} \right )}, M_2 = \frac{\hbar^2}{2 m_0} \sum_{\gamma}^{B}  \frac{2p_{S_x\gamma}^z p_{\gamma S_x}^z }{m_0 \left (E_0-E_{\gamma} \right )}, \\
            M_3 &= \frac{\hbar^2}{2 m_0}\sum_{\gamma}^{B}  \frac{2 p_{1\gamma}^x p_{\gamma 1}^x }{m_0 \left (E_0-E_{\gamma} \right )}, N_1 = \frac{\hbar^2}{m_0^2} \sum_{\gamma}^{B} p_{S_x\gamma}^x p_{S_y\gamma}^y + p_{S_x\gamma}^y p_{S_y\gamma}^x,
        \end{split}
        \end{equation}
where the relation between band parameters ($L_1, M_1, ...$) and inverse effective mass parameters ($A_1, A_2, ...)$ are the same as in Ref.~\cite{chuang1996k}, with $E_0$ is the energy of the band of interest and $E_{\gamma}$ is the band energy of the distant band. Note that, due to the inversion symmetry, there are no coupling terms between CB and VB, unlike in the case of wurtzite materials. Because of  the hexagonal symmetry we can write $N_1 = L_1 - M_1$ and $N_2 = 0$ for our case. In the $k_z$ direction, since the lowest conduction band does not couple to any other band, $A_{c1\parallel} = \hbar^2/(2 m_0)$. Using the bases we have introduced in Table~\ref{table:irreps}, and following the notation of Ref.~\cite{chuang1996k} we can write the Eq.~\eqref{eq:kp2inv} as,
		\begin{eqnarray}
		\label{eq:kp2corrected}
		H_{k \cdot p}^{(2)} =
		\begin{pmatrix}		
                C_2  & 0  &0  &0  &0 & 0 & 0 &0 &0 &0 \\ 
                0 &C_1 &0  &0  &0 & 0 & 0 &0 & 0 &0 \\ 
                0 &0  &\lambda + \alpha  & -K^*  &-T^* & 0 & 0 & 0 & 0 & 0\\ 
                0 &0  &-K & \lambda + \alpha &T &0 &0 &0 &0 &0\\ 
                0 &0  &-T &T^* & \lambda &0 &0 &0 &0 &0 \\
                0 & 0  &0  &0  &0 & C_2 & 0 &0 &0 &0 \\
                0 & 0  &0  &0  &0 & 0 & C_1 &0 &0 &0 \\
                0 &0  &0  & 0  & 0 & 0 & 0 & \lambda + \alpha & -K & T\\ 
                0 &0  &0  & 0  & 0 & 0 & 0 & -K^* & \lambda + \alpha & -T^*\\
                0 &0  &0  & 0  & 0 & 0 & 0 & T^* & -T & \lambda\\ 
		\end{pmatrix}.
		\end{eqnarray}
with the definitions,
    \begin{equation}
        \begin{split}
            C_2 &=  A_{c2\perp}(k_x^2+k_y^2) + A_{c2\parallel}k_z^2, C_1 = A_{c1\perp}(k_x^2+k_y^2) + A_{c1\parallel}k_z^2 , \\
            \lambda & = A_1k_z^2 + A_2 \left (k_x^2 + k_y^2\right), \alpha =  A_3k_z^2 + A_4 \left (k_x^2 + k_y^2\right) , \\
            K &= A_5k_+^2, T = A_6\left (k_x + ik_y)\right)k_z .
        \end{split}
    \end{equation}

\section{Optical Selection Rules for 2H-Ge}
\label{appx:optical_def}
Here we present the optical transition rules for the hexagonal germanium. Table~\ref{table:opt1} shows the general dipole transition rules for the VB's (first row) and CB's (first column). Table~\ref{table:CB} shows the possible transitions between CB and VB, VB-1 and VB-2 with both SOC on and off and similarly Table~\ref{table:CB+1} shows the transitions to the CB+1. Similar analysis has been done by Ref.~\cite{tronc1999optical} for the wurtzite structure. We should point out that CB+1 of the lonsdaleite structure and CB of the wurtzite structure behaves exactly same in terms of allowed optical transitions, for with and without SOC. CB ($\Gamma_8^-$) of the lonsdaleite, however, does not exist in the wurtzite. Thus, Table~\ref{table:CB} is lonsdaleite spesific allowed optical transitions.
\begin{table}[h!]
\centering
\caption{Selection rules for the direct optical transitions with and without SOC for the $D_{6h}$ point group. For the irreps, the parenthesis is used to describe the single group representations of the double groups. For the optical transitions, parentheses (brackets) used when the SOC is (is not) taken into account. Capital letters are used when the transition is allowed  with SOC and without SOC. Also it should be noted that, we don't specify how the irreps change under the inversion symmetry here. Hence, the table should only be used when irreps in the rows and columns are in opposite parity.}
\label{table:opt1}
\begin{tabular}{||c c c c c c c||} 
 \hline
 - & $\Gamma_7(\Gamma_1)$ & $\Gamma_7(\Gamma_2)$ & $\Gamma_7(\Gamma_5)$ & $\Gamma_8(\Gamma_3)$ & $\Gamma_8(\Gamma_4)$ & $\Gamma_8(\Gamma_6)$\\ [0.5ex] 
 \hline\hline
 $\Gamma_7(\Gamma_1)$ & (x,y,z) & (x,y),Z & X,Y,(z) & - & - & - \\ 
 $\Gamma_7(\Gamma_2)$ & (x,y),Z & (x,y,z) & X,Y,(z) & - & - & -\\
 $\Gamma_7(\Gamma_5)$ & X,Y,(z) & (X,Y),z & Z,(x,y) & - & - & [x,y]\\
 $\Gamma_9(\Gamma_5)$ & X,Y & X,Y & [z],(x,y) & (x,y) & (x,y) & X,Y \\
 $\Gamma_9(\Gamma_6)$ & (x,y) & (x,y) & X,Y & X,Y & X,Y & [z],(x,y)\\ [1ex] 
 \hline
\end{tabular}
\label{table:1}
\end{table}

\begin{table}[h!]
\centering
\caption{Selection rules for the lowest conduction band. Transitions from the top three valance band to lowest conduction band is forbidden when SOC is turned off, and dipole allowed when SOC is on. Parenthesis is used to describe the irrep when SOC is neglected.}
\label{table:CB}
 \begin{tabular}{||c c c c||} 
 \hline
 Transitions (CB) &  A : $(\Gamma_8 (\Gamma_3)\leftarrow \Gamma_9 (\Gamma_5$), & B : $\Gamma_8 (\Gamma_3)\leftarrow \Gamma_7 (\Gamma_5$), & C : $(\Gamma_8 (\Gamma_3)\leftarrow \Gamma_7 (\Gamma_1$) \\ [0.5ex] 
 \hline\hline
 \textbf{Neglecting spin-orbit} & - & - & - \\ 
 \textbf{With spin-orbit} & $\Gamma_5(x,y) + \Gamma_6$ & $\Gamma_3 + \Gamma_4  + \Gamma_6$ & $\Gamma_3 + \Gamma_4 + \Gamma_6$ \\ [1ex] 
 \hline
 \end{tabular}
\end{table}

\begin{table}[h!]
\centering
\caption{Selection rules for the second conduction band. For the A and B transitions, optical activity is fully allowed in the x-y directions, even without SOC is on but no optical activity in the z direction. For the C transition, z direction is optically active when SOC is off, and all directions are active when SOC is on.}
\label{table:CB+1}
 \begin{tabular}{||c c c c||} 
 \hline
 Transitions (CB+1) & A : $\Gamma_7 (\Gamma_2)\leftarrow \Gamma_9 (\Gamma_5$), & B : $\Gamma_7 (\Gamma_2)\leftarrow \Gamma_7 (\Gamma_5$), & C : $\Gamma_7 (\Gamma_2)\leftarrow \Gamma_7 (\Gamma_1$) \\ [0.5ex] 
 \hline\hline
 \textbf{Neglecting spin-orbit} & $\Gamma_5(x,y)$ & $\Gamma_5(x,y)$ & $\Gamma_2(z)$ \\ 
 \textbf{With spin-orbit} & $\Gamma_5(x,y) + \Gamma_6$ & $\Gamma_1 + \Gamma_2(z) + \Gamma_5(x,y)$ & $\Gamma_1 + \Gamma_2(z) + \Gamma_5(x,y)$ \\ [1ex] 
 \hline
 \end{tabular}
\end{table}

\section{Twelve-band model at the $\Gamma$ point}
\label{appx:11x11}

In this Appendix, we provide the twelve-band (without spin) $\mathbf{k\cdot p}$ model that is mentioned in the Sec.~\ref{sec:Fit}. This is and extended version of the $5\times 5$ spinless model introduced in Sec.\ref{sec:kp}. The basis functions we use are the same as in Table~\ref{table:irreps}, with the addition of CB+6 and CB+5 that transform as $\Gamma_{6}^-$, CB+4 as $\Gamma_{2}^-$, CB+3 and CB+2 as $\Gamma_{5}^-$ for the conduction bands.  Similarly,  for the valance bands we have VB-3 and VB-4, which transform as $\Gamma_{6}^+$ of the $D_{6h}$ point group.

\begin{table*}[h!]
 \centering
 \caption{$12\times 12\, \mathbf{k.p}$ matrix elements at the $\Gamma$ point. $\mathbf{H}_{k.p}^{(1)}$ represents the first order $\mathbf{k.p}$ terms that has been found in Eq.~\eqref{A2}.}
 \label{table:12x12}
 \begin{NiceTabular}{c c c c c c c c c c c c c}[first-row]
   $H_{k.p}$ & CB+6 & CB+5 & CB+4 & CB+3 & CB+2 & CB+1 & CB & VB & VB-1 & VB-2 & VB-3 & VB-4 \\ \hline
   CB+6 & 0 & 0 & 0 & 0 & 0 & 0 & 0 & 0 & $\gamma_1k_+$ & 0 & $\gamma_2k_z$ & 0 \\
   CB+5 & 0 & 0 & 0 & 0 & 0 & 0 & 0 & $\gamma_3 k_-$ & 0 & 0 & 0 & $\gamma_4 k_z$ \\
   CB+4 & 0 & 0 & 0 & 0 & 0 & 0 & 0 & $\gamma_5 k_+$ & $\gamma_6k_-$ & $\gamma_7k_z$ & 0 & 0 \\
   CB+3 & 0 & 0 & 0 & 0 & 0 & 0 & 0 & 0 & $\gamma_8 k_z$ & $\gamma_{9} k_+$ & $\gamma_{10} k_-$ & 0 \\
   CB+2 & 0 & 0 & 0 & 0 & 0 & 0 & 0 & $\gamma_{11} k_z$ & 0 & $\gamma_{12} k_-$ & 0 & $\gamma_{13} k_+$ \\
   CB+1 & 0 & 0 & 0 & 0 & 0 & \Block{5-5}{\huge $H_{k.p}^{(1)}$} & & & & & 0 & 0 \\
   CB & 0 & 0 & 0 & 0 & 0 && &&&&$\gamma_{14}k_+$ & $\gamma_{15}k_-$\\
   VB & 0 & $\gamma_3^{*}k_+$ & $\gamma_5^{*}k_-$ & 0 & $\gamma_{11}^{*}k_z$ && &&&&0&0 \\
   VB-1 & $\gamma_1^{*}k_-$ & 0 & $\gamma_6^{*}k_+$ & $\gamma_8^{*}k_z$ & 0 && &&&&0&0 \\
   VB-2 & 0 & 0 & $\gamma_7^{*}k_z$ & $\gamma_9^{*}k_-$ & $\gamma_{12}^{*}k_+$  &&&&&&0&0 \\
   VB-3 & $\gamma_2^{*}k_z$ & 0 & 0 & $\gamma_{10}^{*}k_+$ & 0 & 0 & $\gamma_{14}^{*}k_-$ & 0 & 0 & 0 & 0 & 0\\
   VB-4 & 0 & $\gamma_4^{*}k_z$ & 0 & 0 & $\gamma_{13}^{*}k_-$ & 0 & $\gamma_{15}^{*}k_+$  & 0 & 0 & 0 & 0 & 0 \\ \hline
   \CodeAfter
  \SubMatrix[{6-7}{10-11}][extra-height=-3pt,xshift=-3mm]
 \end{NiceTabular}
 \label{tbl:kp-nine-band-G}
\end{table*}
\end{widetext}
\bibliography{references}
\end{document}